\documentclass[onecolumn,prd,showpacs,preprintnumbers,floatfix,article,nofootinbib]{revtex4}
\usepackage{dcolumn}
\usepackage{bm}
\usepackage{longtable}
\usepackage{graphicx,epsfig,latexsym,amssymb}
\usepackage{multirow,amsmath,array,booktabs,color}
\usepackage[section]{placeins}
\usepackage{makecell}
\usepackage{array}

\begin{document}
	\title{Revisit prompt $J/\psi$ production in associated with Higgs Boson via gluon fusion at the LHC}

	\author{Xue-An Pan$^{1}$~}
	\author{Zhong-Ming Niu$^{1}$~}
	\author{Gang Li$^{1}$~}
	\author{Yu Zhang$^{2,1}$~}\email{dayu@ahu.edu.cn}
	\author{Mao Song$^{1}$~}
	\author{Jian-You Guo$^{1}$~}

	\affiliation{$^1$ School of Physics and Materials Science, Anhui University, Hefei, Anhui 230039, People's Republic of China}
	\affiliation{$^2$ Institutes of Physical Science and Information Technology, Anhui University, Hefei, Anhui 230039, People's Republic of China }

	\date{\today }
	\begin{abstract}
The production of charmonium associated with Higgs boson via gluon fusion has been investigated in Ref.\cite{Kniehl:2002wd}, in which they considered the contribution of  final Higgs boson  radiation off the charm quark at tree level and found that this process is to be far too rare to be observable in any of the considered experiments. In this paper, the production of prompt $J/\psi$ associated with Higgs boson via gluon fusion at the 14 TeV LHC within the factorization formalism of NRQCD is revisited.
After considering the contribution from the final Higgs boson radiation off the top quark in the loop,
which is {more than} three orders of magnitudes over the charm quark at tree level,
the production of prompt $J/\psi$ associated with Higgs boson has great potential to be detected.
The prompt $J/\psi$ production includes the direct production and indirect production via radiative or hadronic decays of high excited charmonium states. For the direct $J/\psi + H$ production via gluon fusion loop-induced, the ${}^{3}S^{(8)}_1$ Fock state gives dominant contribution to the cross section, which is about 95\% to the total direct production. The indirect contribution  via loop-induced is appreciable, since the summation of which from $\psi(2S) + H$, $\chi_{c1} + H$ and $\chi_{c2} + H$ is about $34\%$ to the total cross section of prompt $J/\psi + H$. While the indirect contribution from $\chi_{c0} + H$ is tiny, which can be neglected. With the great potential to be detected, prompt $J/\psi$ production in associated with Higgs boson can help us to further understand the mechanism of colour-octet, as well as  can be useful to further investigate the coupling of the Higgs boson and fermion.
	\end{abstract}
	
	\pacs{12.38.Bx, 13.85.Fb, 14.40.Lb, 14.80.Cp}
	\maketitle
	
	\section{Introduction}
	Since the discovery of the first heavy charmonium $J/\psi$ in 1974\cite{Aubert:1974js,Augustin:1974xw}, the production of heavy quarkonium has been an important investigating topic in hadron physics. An increasing number of experiments have studied the production process of heavy quarkonium in detail. For example, the inclusive charmonium production  has been measured by BaBar\cite{Aubert:2001pd} and Belle\cite{Abe:2001za}, the photoproduction of $J/\psi$ has been investigated by H1 \cite{Aid:1996dn,Aid:1996ee,Adloff:2000vm,Aktas:2005xu,Alexa:2013xxa} and ZEUS \cite{Chekanov:2002xi}, and the hadronic and polarization productions of heavy quarkonium have been studied at hadron colliders \cite{Khachatryan:2010zg,Aaij:2014qea,CDF:2011ag,Abulencia:2007us,Aaij:2013nlm}. In addition, {production processes of the double charmonium have been detected, such as $J/\psi + J/\psi$\cite{Abe:2004ww,Aaij:2011yc,Khachatryan:2014iia,Abazov:2014qba}, $J/\psi + \chi_{c0}$\cite{Aubert:2005tj} and $J/\psi + \eta_c$\cite{Aubert:2005tj,Abe:2002rb}. In order to explain the production and decay of the heavy quarkonium, the colour-evaporation model (CEM)\cite{Barger:1979js,Barger:1980mg}, colour-singlet model (CSM)\cite{Ablikim:2020lpe,Berger:1980ni} and nonrelativistic quantum chromodynamics (NRQCD)\cite{Bodwin:1994jh} have been proposed. {{At present}}, NRQCD is more widely accepted and applied to the production and decay of heavy quarkonium. In NRQCD factorization, the production and decay of the heavy quarkonium are factorized into the short-distance coefficients (SDCs) and long distance matrix elements (LDMEs). The SDCs can be perturbatively calculated by using the expansion of the strong-coupling constant $\alpha_s$, and the LDMEs are process-independent and universal, governed by nonperturbative QCD dynamics, and can be extracted from experiments. NRQCD has achieved remarkable success in explaining the puzzle of $J/\psi$ and $\psi(2S)$ surplus production at the Tevatron\cite{Braaten:1994vv,Cho:1995vh}, subsequently it was used extensively to explain the production and decay of heavy quarkonium, and attained {{many phenomenological successes}} (see reviews \cite{Lansberg:2006dh,Lansberg:2019adr,Usachov:2019czc} for details). However, there still exist some problems in NRQCD, such as it cann't provide a full description for the data of the double  $J/\psi$ measured by the CMS\cite{Sun:2014gca,Lansberg:2013qka,Khachatryan:2014iia}, and the universality of LDMEs\cite{Ma:2010yw,Zhang:2009ym} etc., which all indicate that the NRQCD needs further investigating.
	
Due to the clean signature, the associated productions of heavy quarkonium and standard model (SM) bosons are  ideal channels to test the NRQCD, which have been attracted more and more attention. {{The hadronic productions  of heavy quarkonium associated with $\gamma, W^{\pm}, Z$ and Higgs boson have been studied at  leading order (LO) within  NRQCD\cite{Roy:1994vb,Kim:1996bb,Braaten:1998th,Kniehl:2002wd}, and the next-to-leading order (NLO) QCD corrections for the productions of heavy quarkonium associated with $\gamma, W^{\pm}, Z$  have been calculated
in Refs.\cite{Li:2008ym,Li:2010hc,Mao:2011kf,Gang:2012ww,Mao:2013xpa}. Furthermore, for the processes of $J/\psi + W^{\pm}$ and $J/\psi + Z$, the contributions from Single Parton Scattering (SPS) and Double Parton Scattering (DPS) have been investigated \cite{Lansberg:2016rcx,Lansberg:2017chq}}}. In experimental, the productions of $J/\psi + Z$,  $J/\psi + W^{\pm}$ at hadron colliders were also detected \cite{Aad:2014rua,Aad:2014kba,Aaboud:2019wfr}. The production rate of $J/\psi + W^{\pm}$ measured by ATLAS \cite{Aad:2014rua} is an order of magnitude larger than SPS predictions from NRQCD \cite{Li:2010hc,Lansberg:2013wva}, and the gap can be filled by considering the contribution from DPS\cite{Lansberg:2017chq}. {{As for  $J/\psi + Z$}}, even if taking the contribution from NLO CEM SPS + DPS\cite{Lansberg:2016rcx} into account, the behavior of the transverse momentum distribution measured by ATLAS \cite{Aad:2014kba} exhibits slight different  with the theoretical prediction. As regards the production of heavy quarkonium associated with Higgs boson, there is no positive news from both phenomenology and experiment.

In Ref. \cite{Kniehl:2002wd}, Kniehl  \emph{et al.} estimated the production of the charmonium associated with Higgs boson at LO based on NRQCD at $e^{+}e^{-},e^{\pm}p, p\bar{p}$ and $pp$ colliders for the first time. They  considered the contribution from the partonic processes $g + g \rightarrow \mathcal{Q} + H$ and $q + \bar{q} \rightarrow \mathcal{Q} + H$, where the $\mathcal{Q}$ is charmonium, $q$ is the light quark, and $H$ is the Higgs boson, and found that the process of heavy quarkonium associated with Higgs boson is to be far too rare to be observable in any of the considered experiments. { We repeat
the calculation in Ref. \cite{Kniehl:2002wd} for the direct production of $J/\psi +H$ and $\chi_{cJ} + H$, and obtain consistent results of the transverse momentum and rapidity distribution.
Adopting the same input parameters, we obtain the total cross section of $J/\psi$ production in associated with Higgs boson via gluon fusion at the 14 TeV LHC  is about 0.02 fb, which is agree with the results shown in their another work \cite{Kniehl:2004fa}.
One can see that, in this case it is difficult to detect this process at the LHC.  We consider the contribution from the final Higgs boson radiation off top quark via loop-induced, and find that the production of heavy charmonium associated with Higgs boson can be potentially detected at the LHC
\footnote{ Note that
final Higgs boson can also radiate off bottom quark via loop-induced. The contribution of the bottom-loop
is about one order of magnitude large than that considered in Ref.\cite{Kniehl:2002wd}, but nearly three orders of magnitude smaller than that of
top-loop. Therefore, we don't consider it in this work.}.}

	In this work, we will revisit the production of prompt $J/\psi$ associated with Higgs boson via gluon fusion at the LHC within the NRQCD factorization formalism. The final Higgs boson can produce from two sources, including radiation off the charm quark (lablled as S1 in the following) and off the top quark via loop-induced (lablled as S2).	Noted that S1 has been considered by Kniehl \emph{et al.} in  Ref.\cite{Kniehl:2002wd}, and S2 will be considered for the first time in this work. Since the production rate from the partonic process $q + \bar{q} \rightarrow \mathcal{Q} + H$ is tiny, we
	will not consider it. Besides the direct production, prompt $J/\psi$ candidates that can also be produced indirectly via radiative or hadronic decays of heavier charmonium states, such as
	$\chi_{cJ} \rightarrow J/\psi + \gamma\, (J=0,1,2)$ and $\psi(2S) \rightarrow J/\psi + X$.
	We find that the contribution of S2 is {more than} three orders of magnitudes larger than S1, and the $^{3}S^{(8)}_1 + H$ production channel plays an important role in  $J/\psi + H$ production via $gg$ fusion  loop-induced.
	With the detectable potential at the LHC, this process can be useful to test the heavy quarkonium
	production mechanism and investigate the coupling of Higgs and fermion.

	\section{the details of the calculation framework}
	In this section, we introduce the detail of the calculation for prompt $J/\psi$ production in associated with $H$ via gluon fusion, $p p \rightarrow g g \rightarrow J/\psi + H$, at the LHC in NRQCD. The total cross section of prompt $J/\psi+H$ production can be expressed as follow
	\begin{eqnarray}
	\sigma^{\mathrm{prompt}}( \ J/\psi + H) &=& \sigma^{\mathrm{direct}}(J/\psi + H) + \sigma^{\mathrm{indirect}}(\mathrm{From}\ \psi(2S) + H) + \sigma^{\mathrm{indirect}}(\mathrm{From}\ \chi_{cJ} + H) \notag \\
	&=& \sigma(p p \rightarrow g g \rightarrow J/\psi + H) +  \sigma(p p \rightarrow g g \rightarrow \psi(2S) + H)\times \mathrm{Br}(\psi(2S) \rightarrow J/\psi + X) +  \notag \\
	&&\sum_{J=0}^2\sigma(p p \rightarrow g g \rightarrow \chi_{cJ} + H)\times\mathrm{Br}(\chi_{cJ} \rightarrow J/\psi + \gamma),
	\end{eqnarray}
	where the $\sigma^{\mathrm{direct}}(J/\psi + H)$ is the cross section of the direct contribution from $J/\psi + H$ production; $\sigma^{\mathrm{indirect}}(\mathrm{From}\ \psi(2S)/\chi_{cJ} + H)$  denotes the indirect contribution due to hadronic/radiative decays of $\psi(2S)/\chi_{cJ}$ from $\psi(2S)/\chi_{cJ} +H$ production.  $\mathrm{Br}(\psi(2S) \rightarrow J/\psi + X)$ and $\mathrm{Br}(\chi_{cJ} \rightarrow J/\psi + \gamma)$ respectively represent the branching ratio of $\psi(2S)$ and $\chi_{cJ}$ decaying into $J/\psi$, which can be obtained from \cite{Tanabashi:2018oca} as
	\begin{eqnarray}
	\mathrm{Br}(\psi(2S) \rightarrow J/\psi + X) &=& (61.4 \pm 0.6)\%,         \notag    \\
	\mathrm{Br}(\chi_{c0} \rightarrow J/\psi + \gamma) &=& (1.40 \pm 0.05)\%,  \notag    \\
	\mathrm{Br}(\chi_{c1} \rightarrow J/\psi + \gamma) &=& (34.3 \pm 1.0)\%,             \\
	\mathrm{Br}(\chi_{c2} \rightarrow J/\psi + \gamma) &=& (19.0 \pm 0.5)\%.    \notag
	\end{eqnarray}
	
	$\sigma(p p \rightarrow g g \rightarrow \mathcal{Q}+H)$ denotes the cross section for the production of charmonium $\mathcal{Q}$ associated with $H$ via gluon fusion, which can be factorized as
	\begin{eqnarray}
	\sigma(p p \rightarrow g g \rightarrow \mathcal{Q}+H)=\int d x_{1} d x_{2} f_{g / A}\left(x_{1}, \mu_{f}\right) f_{g / B}\left(x_{2}, \mu_{f}\right) \notag \\
	\times \sum_{n}\left\langle \mathcal{O}^{\mathcal{Q}}[n]\right\rangle \hat{\sigma}(g + g \rightarrow c\overline{c}[n]+H).
	\end{eqnarray}
	Here $f_{g/A,B}\left(x_{1,2}, \mu_{f}\right)$ is the distribution function of the gluon radiated by proton $A$ or $B$; $x_{1/2}$ denotes the momentum fraction of the proton $A/B$ momentum carried by the gluon; the summation is taken over the possible intermediate Fock states $n =^{2S+1}L^{(1,8)}_J$ of heavy $c\bar{c}$ pair; $\left\langle \mathcal{O}^{\mathcal{Q}}[n]\right\rangle$ is the LDME, which describes the process for the Fock state $c\overline{c}[n]$ evolving into an observable physical state $\mathcal{Q}$ by radiating soft gluons, with $\mathcal{Q} = J/\psi, \psi(2S), \chi_{cJ}(J=0,1,2)$ in our calculation. The relations between the LDMEs of various Fock states are
	\begin{eqnarray}
	\left\langle O^{J / \psi, \psi(2 S)}\left[{ }^{3} P_{J}^{(8)}\right]\right\rangle &=&(2 J+1)\left\langle O^{J / \psi, \psi(2 S)}\left[{ }^{3} P_{0}^{(8)}\right]\right\rangle, \notag \\
	\left\langle O^{\chi_{c J}}\left[{ }^{3} S_{1}^{(8)}\right]\right\rangle &=& (2 J+1)\left\langle O^{\chi_{c0}}\left[{ }^{3} S_{1}^{(8)}\right]\right\rangle, \notag \\
	\left\langle O^{\chi_{c J}}\left[{ }^{3} P_{J}^{(1)}\right]\right\rangle &=& (2 J+1)\left\langle O^{\chi_{c0}}\left[{ }^{3} P_{0}^{(1)}\right]\right\rangle.
	\label{LDME}
	\end{eqnarray}
	
	$\hat{\sigma}(g + g \rightarrow c\overline{c}[n]+H)$ describes the short-distance cross section for the partonic process $g(p_1) + g(p_2) \rightarrow c\overline{c}[n](p_3) + H(p_4)$ via gluon fusion, which can be expressed as
	\begin{eqnarray}
	\hat{\sigma}(g + g \rightarrow c\overline{c}[n] + H)=\frac{1}{16 \pi \hat{s} N_{col} N_{pol}} \int_{\hat{t}_{\min }}^{\hat{t}_{\max }} d \hat{t} \overline{\sum}\left|A_{S, L}\right|^{2}.
	\end{eqnarray}
	Here $\hat{s}$ and $\hat{t}$ are Mandelstam variables, which are defined as $\hat{s}=\left(p_{1}+p_{2}\right)^{2}$ and $\hat{t}=\left(p_{1}-p_{3}\right)^{2}$, respectively. $N_{col}$  ($ N_{pol}$) refers to the colour (polarization) quantum number of Fock states $c\overline{c}[n]$. The summation is taken over the initial state and the final state, and the bar means averaging the colour and spin of the initial parton. $A_{S, L}$ can be obtained by the covariant projection method in Ref.\cite{Petrelli:1997ge}. The polarization summations of the Fock state $c\overline{c}[n]$ are
	\begin{eqnarray}
	\sum_{J_{z}} \varepsilon_{\alpha} \varepsilon_{\alpha^{\prime}}^{*} &=& \Pi_{\alpha \alpha^{\prime}}, \notag \\
	\sum_{J_{z}} \varepsilon_{\alpha \beta}^{0} \varepsilon_{\alpha^{\prime} \beta^{\prime}}^{0*} &=& \frac{1}{3} \Pi_{\alpha \beta} \Pi_{\alpha^{\prime} \beta^{\prime}}, \notag\\
	\sum_{J_{z}} \varepsilon_{\alpha \beta}^{1} \varepsilon_{\alpha^{\prime} \beta^{\prime}}^{1*}&=&\frac{1}{2}\left(\Pi_{\alpha \alpha^{\prime}} \Pi_{\beta \beta^{\prime}}-\Pi_{\alpha \beta^{\prime}} \Pi_{\alpha^{\prime} \beta}\right), \notag \\
	\sum_{J_{z}} \varepsilon_{\alpha \beta}^{2} \varepsilon_{\alpha^{\prime} \beta^{\prime}}^{2*}&=&\frac{1}{2}\left(\Pi_{\alpha \alpha^{\prime}} \Pi_{\beta \beta^{\prime}}+\Pi_{\alpha \beta^{\prime}} \Pi_{\alpha^{\prime} \beta}\right)-\frac{1}{3} \Pi_{\alpha \beta} \Pi_{\alpha^{\prime} \beta^{\prime}}
	\end{eqnarray}
	with
	\begin{eqnarray}
	\Pi_{\alpha \beta}=-g_{\alpha \beta}+\frac{p_{3 \alpha} p_{3 \beta}}{M_{Q}}.
	\end{eqnarray}
	Here $p_3$ ($M_{Q}$) is the momentum (mass) of quarkonium $\mathcal{Q}$.
	
	The polarization summation of the two initial gluons for $g + g \rightarrow c\overline{c}[n]+H$ process is expressed as follows
	\begin{eqnarray}
	-g_{\mu \nu}+\frac{p_{\mu} \eta_{\nu}+p_{\nu} \eta_{\mu}}{p \cdot \eta},
	\end{eqnarray}
	where $p$ is the momentum of initial gluon, and
	$\eta$ is an arbitrary light-like vector with $p \cdot \eta \neq 0$. In our calculation, $\eta$ is taken as the momentum of another initial gluon.

	{The Feynman diagrams for the process of $g + g \rightarrow c\overline{c}[n]+H$ from the contribution of S1 and S2 are generated with FeynArts\cite{Hahn:2000kx}, which are presented in Fig. \ref{fig5} and Fig. \ref{fig1} respectively}, and we further reduce the Feynman amplitudes using FeynCalc\cite{Shtabovenko:2016sxi}, Apart\cite{Feng:2012iq,Feng:2013sba} and FIRE\cite{Smirnov:2008iw,Fleischer:1999hq}. Finally, we use LoopTools and FormCalc\cite{Hahn:1998yk} to implement the numerical calculations.

	\begin{figure}[!htbp]
		\includegraphics[scale=0.8]{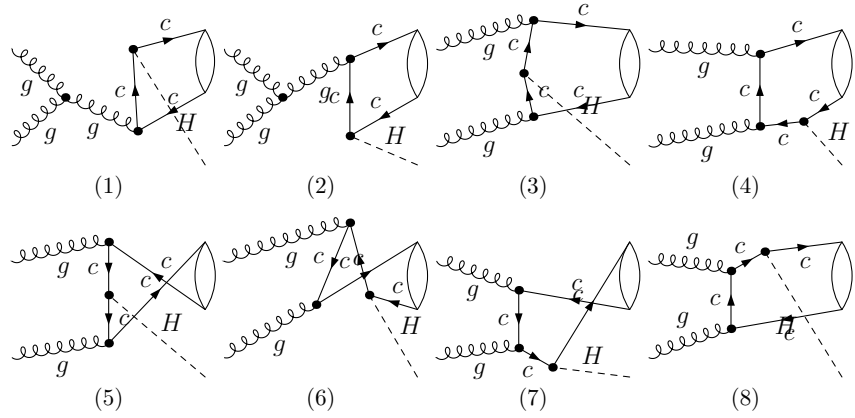}
		\caption{Feynman diagrams for the partonic process $g + g \rightarrow c\overline{c}[n]+H$ via $gg$ fusion  with final Higgs boson radiation off charm quark.}
		\label{fig5}
	\end{figure}

	\begin{figure}[!htbp]
		\includegraphics[scale=0.8]{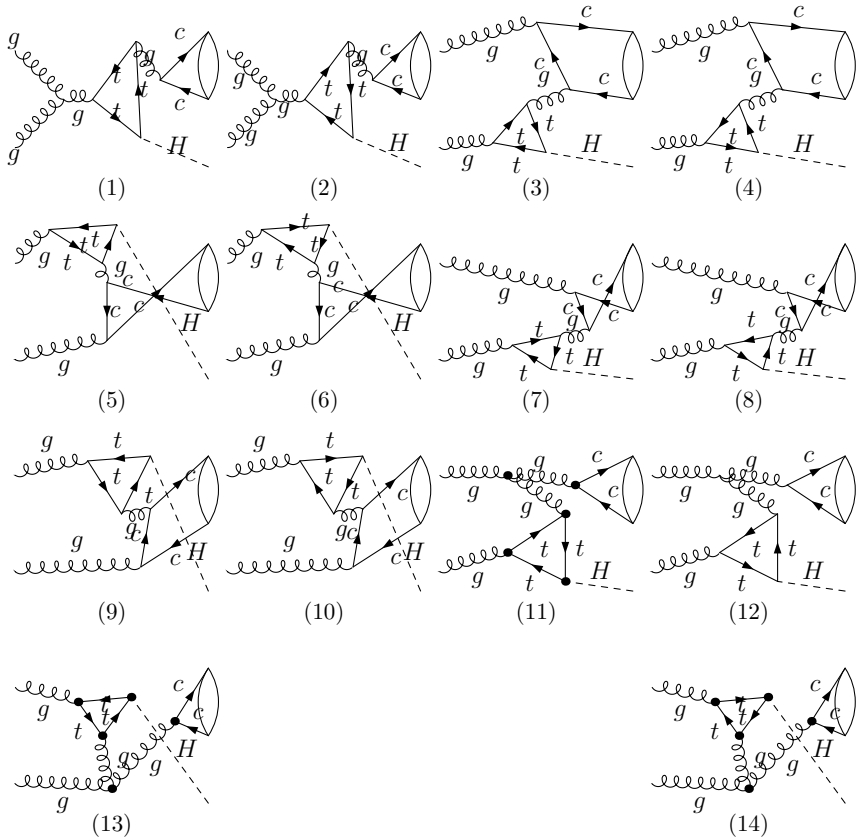}
		\caption{Feynman diagrams for the partonic process $g + g \rightarrow c\overline{c}[n]+H$ via $gg$ fusion with final Higgs boson radiation off top quark in the loop.}
		\label{fig1}
	\end{figure}

	\section{numerical results}

In what follows, we present numerical results for the production of prompt $J/\psi$ associated with Higgs boson via $gg$ fusion from the contribution of S1 and S2 at the LHC with $\sqrt{s}$ = 14 TeV, respectively. The masses of the charm quark, top quark and Higgs boson are taken as $m_c = 1.5\ \rm{GeV}$, $m_t = 172\ \rm{GeV}$, and $m_H = 125\ \rm{GeV}$. The mass of the charmonium $\mathcal{Q}$ is taken as $M_{\mathcal{Q}} = 2 m_c$, and we set cut $p^{\mathcal{Q}}_{T}>3\ \rm{GeV}$ for final charmonium $\mathcal{Q}$, where the $p^{\mathcal{Q}}_{T}$ is the transverse momentum of charmonium $\mathcal{Q}$. In our calculation, the factorization scale and PDFs are chosen as $\mu_f = \sqrt{(p^{\mathcal{Q}}_{T})^{2}+M^{2}_{\mathcal{Q}}}$ and CTEQ6L1\cite{Pumplin:2002vw}, respectively.
	
	The LDMEs of $J/\psi$ is chosen from \cite{Fleming:1997fq,Kniehl:2001tk} as
	\begin{eqnarray}
	\left\langle\mathcal{O}^{J / \psi}\left[{ }^{1} S_{0}^{(8)}\right]\right\rangle &=& 1 \times 10^{-2}\ \mathrm{GeV}^{3}, \notag \\
	\left\langle\mathcal{O}^{J / \psi}\left[{ }^{3} P_{0}^{(8)}\right]\right\rangle &=& 11.25 \times 10^{-3}\ \mathrm{GeV}^{5},  \\
	\left\langle\mathcal{O}^{J / \psi}\left[{ }^{3} S_{1}^{(8)}\right]\right\rangle &=& 1.12 \times 10^{-2}\ \mathrm{GeV}^{3}. \notag
	\end{eqnarray}
	The LDMEs of $\psi(2S)$ and $\chi_{c0}$ are taken from \cite{Kniehl:2006qq,Eichten:1995ch,Ma:2010vd} as
	\begin{eqnarray}
	\left\langle\mathcal{O}^{\psi^{\prime}}\left[{ }^{1} S_{0}^{(8)}\right]\right\rangle &=&5 \times 10^{-3}\ \mathrm{GeV}^{3}, \notag \\
	\left\langle{\mathcal{O}}^{\psi^{\prime}}\left[{ }^{3} P_{0}^{(8)}\right]\right\rangle &=& 3.214 \times 10^{-3}\ \mathrm{GeV}^{5}, \\
	\left\langle\mathcal{O}^{\psi^{\prime}}\left[{ }^{3} S_{1}^{(8)}\right]\right\rangle &=&2 \times 10^{-3}\ \mathrm{GeV}^{3}, \notag
	\end{eqnarray}
	and
	\begin{eqnarray}
	\left\langle\mathcal{O}^{\chi_{c 0}}\left[{ }^{3} S_{1}^{(8)}\right]\right\rangle &=& 2.2 \times 10^{-3}\ \mathrm{GeV}^{3}, \notag \\
	\left\langle\mathcal{O}^{\chi_{c 0}}\left[{ }^{3} P_{0}^{(1)}\right]\right\rangle &=& \frac{3 N_{c}}{2 \pi} \times 0.075\ \mathrm{GeV}^{5},
	\end{eqnarray}
	with $N_c = 3$. In accordance with Eq.~(\ref{LDME}), we can acquire the LDMEs of $\chi_{c1}$ and $\chi_{c2}$. In the calculation, we have considered the conventions of  different theories for LDMEs of the colour singlet states.

In Tab. \ref{Table1}, {we present the cross section of the charmonium $J/\psi$, $\psi(2S)$ and $\chi_{cJ}(J = 0, 1, 2)$ production in associated with the Higgs boson via $gg$ fusion from the contribution of S1 and S2 at the 14 TeV LHC within the NRQCD framework. We can find that the contribution of S2 is {more than} three orders of magnitudes larger than S1 in the production of charmonium associated with Higgs boson.  The Fock state ${}^{3}S^{(8)}_1$ plays a major role in the direct production of charmonium associated with Higgs boson via loop-induced $gg$ fusion}, from which the contribution can account for more than 90\% to the total direct production cross section.
	For the direct production of $J/\psi + H$ and $\chi_{cJ} + H$ via loop-induced, the cross section can reach about 45.34 fb and 81.82 fb respectively, which can provide  abundant and fascinating investigations of phenomenology at the LHC.

\begin{table}[!htbp]
\begin{tabular}{c|c|c|c|c|c|c}
\hline\hline
                & Fock state  & ${ }^{1} S_{0}^{(8)}$ & ${ }^{3} P_{J}^{(8)}$ & ${ }^{3} P_{J}^{(1)}$ & ${ }^{3} S_{1}^{(8)}$ & Total \\
                 \hline
\multirow{5}{*}{S1} & $\sigma\left(p p \rightarrow gg \rightarrow J/\psi +  H\right)$ &  0.97 & 2.64 & $-$ & 4.84 & 8.45 \\ \cline{2-7}
                  & $\sigma\left(p p \rightarrow gg \rightarrow \psi(2S) + H\right)$ & 0.49 & 0.76 & $-$ & 0.80 & 2.05 \\ \cline{2-7}
                  & $\sigma\left(p p \rightarrow gg \rightarrow \chi_{c 0} + H\right)$ &$-$  & $-$ & 0.62 & 0.95 & 1.57 \\ \cline{2-7}
                  & $\sigma\left(p p \rightarrow gg \rightarrow \chi_{c 1} + H\right)$ & $-$  & $-$  &  3.81 & 2.85 &6.66  \\ \cline{2-7}
                  & $\sigma\left(p p \rightarrow gg \rightarrow \chi_{c 2} + H\right)$ & $-$ & $-$ & 9.02 & 4.75  &13.77  \\ \hline
\multirow{5}{*}{S2} & $\sigma\left(p p \rightarrow gg \rightarrow J/\psi +  H\right)$ & 0.38& 0.71 & $-$ & 44.25 & 45.34 \\ \cline{2-7}
                  & $\sigma\left(p p \rightarrow gg \rightarrow \psi(2S) + H\right)$ & 0.19 & 0.20 & $-$ & 7.90  & 8.29  \\ \cline{2-7}
                  & $\sigma\left(p p \rightarrow gg \rightarrow \chi_{c 0} + H\right)$ & $-$  & $-$ & 0.69 & 8.69 & 9.36  \\ \cline{2-7}
                  & $\sigma\left(p p \rightarrow gg \rightarrow \chi_{c 1} + H\right)$ & $-$  & $-$ & 1.75 & 26.08 & 27.82  \\ \cline{2-7}
                  & $\sigma\left(p p \rightarrow gg \rightarrow \chi_{c 2} + H\right)$ & $-$  & $-$ & 1.19 & 43.47 & 44.64  \\ \hline	\hline
\end{tabular}
\caption{The cross section  of charmonium $\mathcal{Q}$ production in associated with Higgs boson via $gg$ fusion from the contrition of S1 (in unit of ab)  and S2 (in unit of fb) at the 14 {TeV} LHC.}
\label{Table1}
\end{table}

	We present the transverse momentum ($p^{\mathcal{Q}}_T$) and rapidity ($y_{\mathcal{Q}}$) distributions of the final charmonium $\mathcal{Q}$ for the process $pp \rightarrow gg \rightarrow \mathcal{Q} + H$ via loop-induced production
	at the 14 TeV LHC in Fig.\ref{fig2}.
	It can be found that, the $p_T$ distributions for  the considered direct $\mathcal{Q}$ production all decrease rapidly with the increment of $p^{\mathcal{Q}}_T$.  The $p_T$ distributions of the direct $\chi_{c2}$ and $J/\psi$ production are almost the same in the whole $p_T$ region, because the Fock state ${}^3S^{(8)}_1$ dominants the contributions and LDME values of the $^{3}S^{(8)}_1$ for $J/\psi$ and $\chi_{c2}$ are quite close.
	Due to same reason, rapidity distributions of $J/\psi$ and $\chi_{c2}$ are also almost consistent.
		
	\begin{figure}[!htbp]
		\vspace*{0cm}
		\hspace*{0cm}
		\includegraphics[width=17cm]{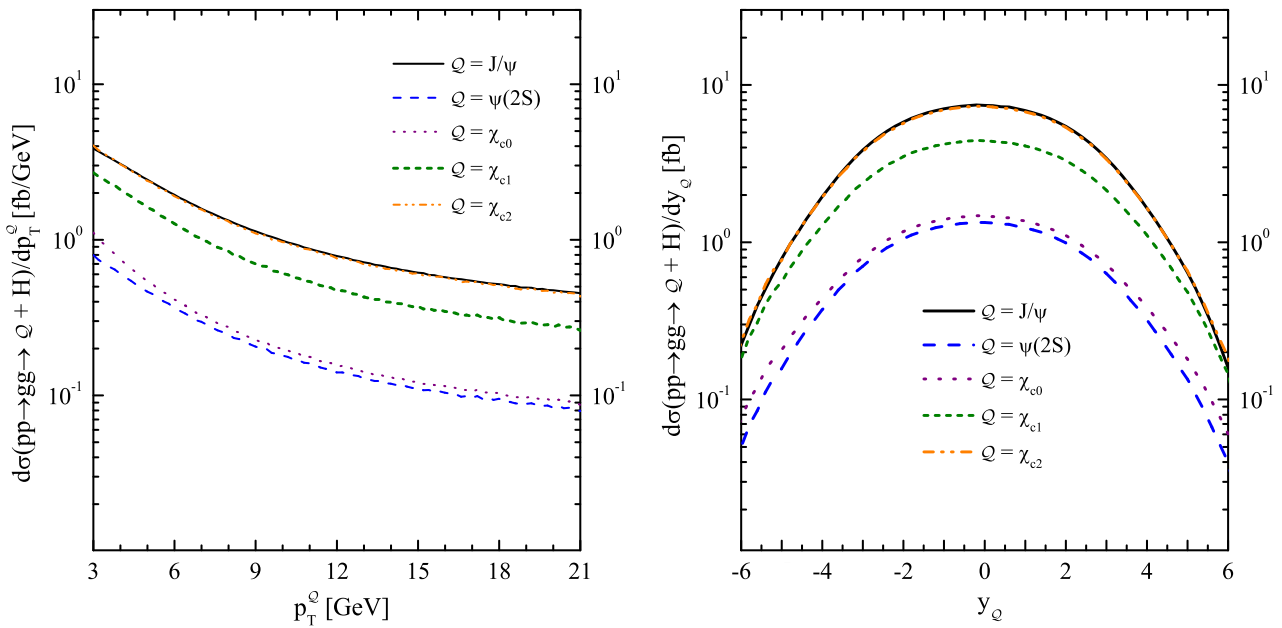}\newline
		\vspace*{0cm}
		\caption{The distributions of $p^{\mathcal{Q}}_T$ (left) and $y_{\mathcal{Q}}$ (right) for the process $pp \rightarrow gg \rightarrow \mathcal{Q} + H$ via loop-induced at the 14 TeV LHC. The black solid, blue dashed, purple dotted, olive short dashed and orange dashed-dotted-dotted lines denote the final charmonium $\mathcal{Q}$ are $J/\psi$, $\psi(2S)$, $\chi_{c0}$, $\chi_{c1}$ and $\chi_{c2}$,respectively.}
		\label{fig2}
	\end{figure}
	
		In order to illustrate the transverse momentum $p_T$ and rapidity $y$ distributions from each Fock state, we take direct $J/\psi$ production in associated with Higgs boson as an example and present the contribution from Fock states $^{1}S^{(8)}_0$, $^{3}P^{(8)}_J$ and $^{3}S^{(8)}_1$ for $J/\psi$ in Fig. \ref{fig3}, separately.
	The production channel of $^{3}S^{(1)}_1 + H$ is forbidden due to charge-conjugation invariance.  We can see that the contribution from each Fock state decreases rapidly with the increment of $p^{J/\psi}_T$, but $^{3}S^{(8)}_1$ decreases slower than $^{1}S^{(8)}_0$ and $^{3}P^{(8)}_J$.
In the range of $ 3\ {\rm GeV} \le \ p^{J/\psi}_T \le  15 \ \rm{GeV}$, the differential cross section of $J/\psi$ associated with Higgs boson production lies in the range of [0.62,3.85] fb/GeV. From Fig. \ref{fig3}, we can find that the contribution from $^{3}S^{(8)}_1$ Fock state is always dominant, especially at large $p_T$  the contribution due to  the other Fock states can be negligible.
	It implies that this process will help to extract the LDMEs of $^{3}S^{(8)}_1$ Fock state if it is detected
	in future experiments.
	
	\begin{figure}[!htbp]
		\vspace*{0cm}
		\hspace*{0cm}
		\includegraphics[width=17cm]{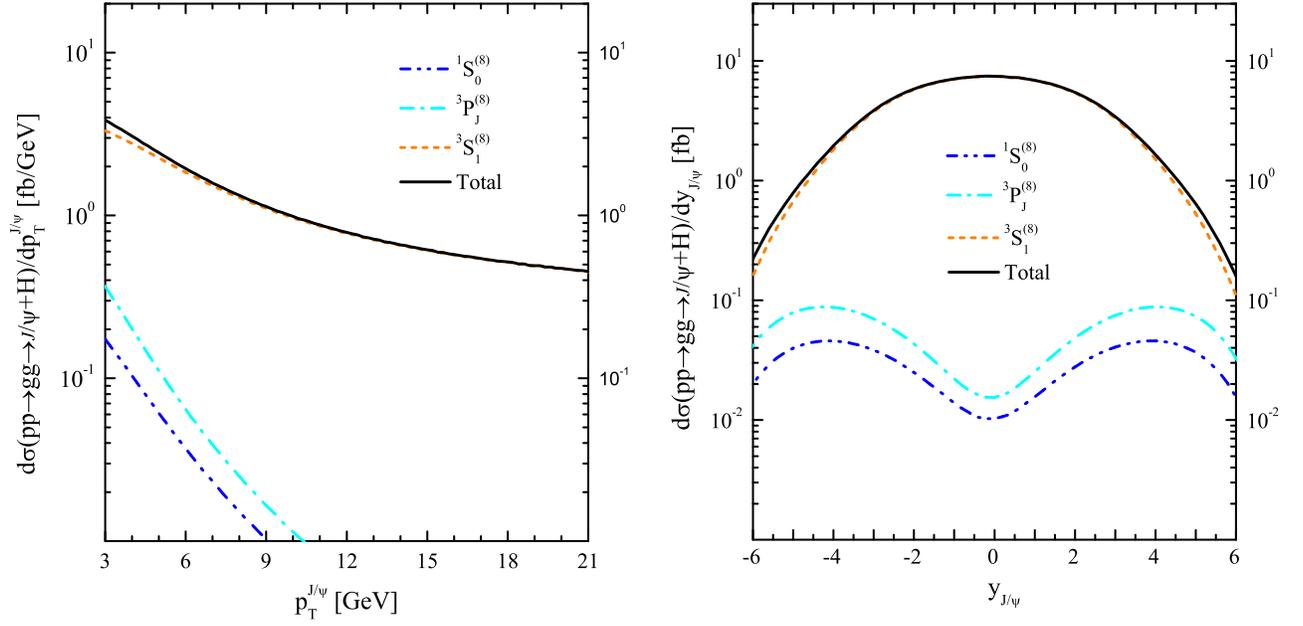}\newline
		\vspace*{0cm}
		\caption{The distributions of $p^{J/\psi}_T$ (left) and $y_{J/\psi}$ (right) for direct production of the $J/\psi$ associated with Higgs boson via loop-induced at the 14 TeV LHC. The blue dashed-dotted-dotted, cyan dashed-dotted, orange dashed and black solid lines are for ${}^1S^{(8)}_0$, ${}^3P^{(8)}_J$, ${}^3S^{(8)}_1$ and total, respectively.}
		\label{fig3}
	\end{figure}

{We respectively present the total cross section for the prompt $J/\psi$ associated with Higgs boson via $gg$ fusion from the contribution of S1 and S2 at the 14 TeV LHC }in Tab. \ref{Table2}. The contributions due to direct and indirect production are also shown respectively. From Tab. \ref{Table2}, {we can see that for S1, even considering the direct and indirect contributions, the cross section of prompt $J/\psi+H$ production is still tiny compared to S2. As for the contribution of S2,} we can find that direct production is dominant, and can reach about 66\% of the prompt production. Indirect production due to excited charmonium decays is not negligible, whose rate can  account for about 34\%. In indirect production, the contributions of $\psi(2S) + H$, $\chi_{c1} + H$ and $\chi_{c2} + H$ are sizable, which are about 22$\%$, 41$\%$ and 36$\%$ of the indirect contribution, respectively. The indirect contribution from $\chi_{c0} + H$ is less than 0.2$\%$ of the prompt production, because the decay branching ratio $\mathrm{Br}(\chi_{c0} \rightarrow J/\psi + \gamma)$ is tiny. Considering the contribution from direct and indirect productions, the total cross section of prompt $J/\psi + H$ production can reach 68.58 fb.
	
	 {
We further consider the final Higgs boson decaying into $b\bar b$. For the $b$-jets, applying the cuts of $p_T^b > 20 \ {\rm GeV}, |\eta_b|< 2.5$, and considering the branching ratio Br$(H \rightarrow b\bar{b}) = 58.14\%$\cite{Dawson:2018dcd} and b-tagging efficiency  $\epsilon_b=77\%$ \cite{Li:2019ipy}, we can get the cross section from S2 as 11.12 (5.7) fb for the direct (indirect) production of $J/\psi$ associated with Higgs boson via $gg$ fusion. Therefore, the cross section of prompt $J/\psi +H$ with Higgs boson decaying into $b\bar b$ can reach about 18 fb even after considering the kinematical constraints on the Higgs in experiments. This implies that this process may be detected in future experiments, which will help us to further understand the colour-octet mechanism. }
	
	\begin{table}[!htbp]
		\begin{tabular}{c| c | c | c | c | c | c | c }
			\hline \hline   Source                        &  From $\psi(2S) + H$ & From $\chi_{c0} + H$ & From $\chi_{c1} + H$ & From $\chi_{c2} + H$ & Indirect & Direct & Prompt   \\
			\hline
			 $\sigma\left(S1\right)$ &       1.26        &        0.02      &       2.28     &        2.62 &    6.18   &  8.45  &   14.63   \\
			\hline
            $\sigma\left(S2\right)$ &       5.09       &        0.13      &        9.54     &        8.48     &   23.24  &   45.34 &   68.58   \\
			\hline \hline
		\end{tabular}
\caption{The cross sections for the prompt $J/\psi$ associated with Higgs boson via $gg$ fusion including the contribution of S1 (in unit of ab) and S2 (in unit of fb) from direct and indirect production at the 14 {TeV} LHC, respectively.}
		\label{Table2}
	\end{table}
	
	\begin{figure}[!htbp]
		\vspace*{0cm}
		\hspace*{0cm}
		\includegraphics[width=17cm]{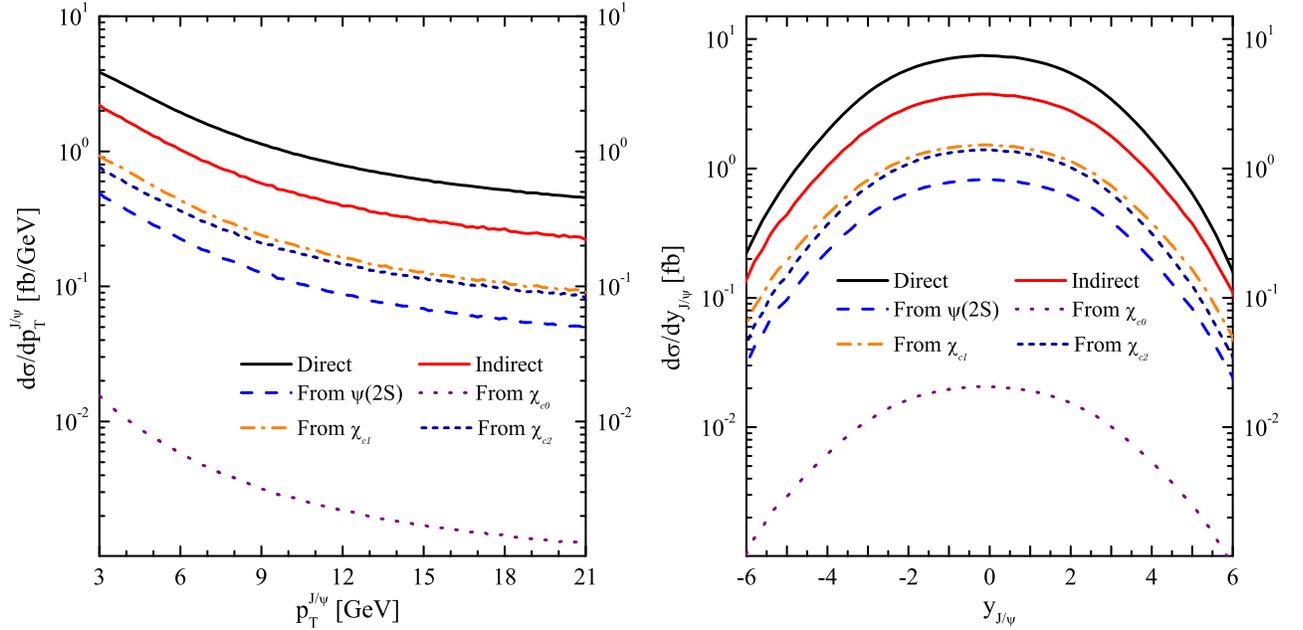}\newline
		\vspace*{0cm}
		\caption{The distributions of $p^{J/\psi}_T$ (left) and $y_{J/\psi}$ (right) for direct and indirect contribution of the $J/\psi$ associated with Higgs boson via loop-induced at the 14 TeV LHC. The black solid, red solid, blue dashed, purple dotted, orange dashed-dotted and royal short dashed lines are contributions of from direct, indirect, $\psi(2S)$, $\chi_{c0}$, $\chi_{c1}$ and $\chi_{c2}$.}
		\label{fig4}
	\end{figure}
	
	The distributions of $p^{J/\psi}_T$ (left) and $y_{J/\psi}$ (right) for direct and indirect productions of the $J/\psi$ associated with Higgs boson via loop-induced at the 14 TeV LHC are shown in Fig \ref{fig4}. It can be seen that the behavior of the $p^{J/\psi}_T$ and $y_{J/\psi}$ distribution for the direct and indirect productions are similar.  The curves of the $J/\psi + H$ direct production are always at the top in the whole plotted region of $p^{J/\psi}_T$ and $y_{J/\psi}$. The indirect contributions from $\psi(2S)$, $\chi_{c1}$ and $\chi_{c2}$ are considerable, while from $\chi_{c0}$ is much less than from other charmonia because of the small  decay branching ratio.
 For the $y_{J/\psi}$ distributions, the convex curves of the direct and indirect contributions reach their peaks when $y_{J/\psi} = 0$.

\begin{table}[!htbp]
		\begin{tabular}{c| c | c | c | c  }
			\hline \hline
                                                         &       Set  1    & Set 2   &  Set 3    & Default set   \\
			\hline
			$<{\cal O}^{J/\psi}[^1S_0^{[8]}]>$ (GeV$^3$) &      $4.35\times10^{-2}$    &  $5.45\times10^{-2}$   &  $0.78\times10^{-2}$     &    $1\times10^{-2}$         \\
			\hline
            $<{\cal O}^{J/\psi}[^3P_0^{[8]}]>$ (GeV$^5$) &       $2.879\times10^{-2}$  &    $3.50\times10^{-2}$   &  $4.3515\times10^{-2}$      &   $1.125\times10^{-2}$          \\
            \hline
            $<{\cal O}^{J/\psi}[^3S_1^{[8]}]>$ (GeV$^3$) &         $4.4\times10^{-3}$    &     $1.4\times10^{-2}$    &    $1.057\times10^{-2}$        &  $1.12\times10^{-2}$                   \\
            \hline
            $\sigma(pp \rightarrow gg \rightarrow J/\psi +  H)({\rm fb})$&  20.85        &      59.59            &  44.81                &       45.34     \\
			\hline \hline
		\end{tabular}
\caption{The cross sections (fb) of the direct $J/\psi$ associated with Higgs boson production via $gg$ fusion from the contribution of S2 at the 14 TeV LHC with four different sets of LDMEs, in which Set 1 and Set 2 are extracted by fitting the $p_T$ distributions from CDF using PDFs for MRST98LO\cite{Braaten:1999qk} and MRS(R2)\cite{Beneke:1996yw}, respectively. And Set 3 is extracted from the hadroproduction of $\eta_c$ at the LHC\cite{Sun:2015hhv}. }
		\label{Table3}
\end{table}

{As we know, in the study of heavy quarkonium production and decay processes, LDMEs is considered universal. The LDMEs of colour octet are extracted from experiments, which lead to the uncertainty. In order to study the uncertainty of the results caused by the LDMEs,  we present the cross sections of the direct $J/\psi$ associated with Higgs boson production via $gg$ fusion from the contribution of S2 at the 14 TeV LHC with other three different sets of LDMEs in Tab. \ref{Table3}. Set 1 and Set 2 are extracted by fitting the $p_T$ distributions from CDF using PDFs for MRST98LO\cite{Braaten:1999qk} and MRS(R2)\cite{Beneke:1996yw}, respectively; Set 3 is extracted from the hadroproduction of $\eta_c$ at the LHC\cite{Sun:2015hhv}. From  Tab. \ref{Table3}, we can see that the cross section of $J/\psi + H$ varies between $46\%-131\%$ compared with the result with the default set of LDMEs. Moreover, we take the production channel of $J/\psi(^3S_1^{(8)})$ as an example to study the uncertainty due to PDFs. In Tab. \ref{Table4}, we respectively show the cross sections of $J/\psi(^3S_1^{(8)}) + H$ via $gg$ fusion from the contribution of S2 at the 14 TeV LHC with other three different PDFs. From Tab. \ref{Table4}, we can find that the results are slightly affected by the different PDFs. In addition to the  LDMEs and PDFs mentioned above, the numerical uncertainty in our calculation is also affected by the factorization scale and quark mass. }

	\begin{table}[!htbp]
		\begin{tabular}{c| c | c | c | c  }
			\hline \hline
                             PDFs                        &      MSHT20lo$\_$as130\cite{Bailey:2020ooq}   & NNPDF30lo$\_$as$\_$0130\cite{Ball:2014uwa}  &   NNPDF31$\_$lo$\_$pch$\_$as$\_$0130\cite{Ball:2017nwa}  & Default   \\
			\hline
			$\sigma(pp \rightarrow gg \rightarrow J/\psi(^3S_1^{(8)}) +  H)({\rm fb})$ & 45.01  &  43.50   &  42.65     &    44.25      \\
			\hline \hline
		\end{tabular}
        \caption{The cross sections of the  $J/\psi(^3S_1^{(8)}) + H$ production channel via $gg$ fusion from the contribution of S2 at the 14 {TeV} LHC with four different PDFs, respectively.}
		\label{Table4}
	\end{table}

	\section{summary and discussion}
	In this paper, we revisit the production of prompt $J/\psi$ associated with Higgs boson via $gg$ fusion include the contribution of S1 and S2 at the 14 TeV LHC within the framework of NRQCD. We consider the direct contribution from $J/\psi + H$ and indirect contribution from $\psi(2S) + H$ and $\chi_{cJ} + H$.  From our calculations, we find that the contribution of S2 is { more than} three orders of magnitudes larger than  S1 for the production  of charmonium associated with Higgs boson. Not like the contribution of S1, which is to be far too rare
	to be observable, the contribution of S2 has great potential to be detected. We not only give the total cross section of prompt $J/\psi + H$, but also present the distribution of transverse momentum and rapidity for the final $J/\psi$ in the contribution of S2. We find that the total cross section of prompt $J/\psi + H$ can reach about 70 fb and the ${ }^{3} S_{1}^{(8)}$ Fock state gives the dominant contribution in all production channels. In the  process of prompt $J/\psi + H$ production via loop-induced, the direct contribution is a majority, while the total indirect contribution is also considerable, accounting for 34$\%$ of the prompt $J/\psi + H$. The productions of $\psi(2S) + H$, $\chi_{c1} + H$ and $\chi_{c2} + H$ provide 22$\%$, 41$\%$ and 36$\%$ of the total indirect contribution, respectively. Since the decay channel of $J/\psi$ is clean, it is possible to detect the production of $J/\psi$ associated with the Higgs boson at the LHC in the future, which will help us to further understand the mechanism of colour-octet, as well as help us to further investigate the coupling of the Higgs boson and fermion.
	
	\section*{Acknowledgments}
	This work was supported in part by the National Natural
	Science Foundation of China (Grants No. 11805001, No. 11935001, No.11875070).

\end{document}